\documentclass[prd,12pt,letterpaper,nofootinbib]{revtex4}
\usepackage{natbib}
\usepackage{amssymb,graphicx}

\newcommand{\Mpc}{\mbox{ Mpc}}

\newcommand{\Mpcinv}{\mbox{ Mpc$^{-1}$}}

\newcommand{\sqm}{\mbox{ m$^2$}}

\newcommand{\kel}{\mbox{ K}}
\newcommand{\mkel}{\mbox{ mK}}

\newcommand{\hr}{\mbox{ hr}}

\newcommand{\MHz}{\mbox{ MHz}}

\newcommand{\bxhi}{\bar{x}_{\rm HI}}
\newcommand{\xhi}{x_{\rm HI}}

\newcommand{\dtb}{\delta T_b}
\newcommand{\bdtb}{\bar{\delta T}_b}

\newcommand{\lya}{Ly$\alpha$ }

\newcommand{\deriv}{{\rm d}}
\newcommand{\bq}{\begin{equation}}
\newcommand{\eq}{\end{equation}}
\newcommand{\bqa}{\begin{eqnarray}}
\newcommand{\eqa}{\end{eqnarray}}
\def\VEV#1{\left\langle #1\right\rangle} 

\begin{document}

\title{Cosmology from the Highly-Redshifted 21 cm Line}

\author{Steven~R.~Furlanetto$^1$, 
Adam~Lidz$^2$, 
Abraham~Loeb$^2$, 
Matthew~McQuinn$^2$, 
Jonathan~R. Pritchard$^2$, 
Paul~R.~Shapiro$^3$,
James~Aguirre$^4$,
Marcelo~A.~Alvarez$^5$,
Donald~C.~Backer$^6$, 
Judd~D.~Bowman$^7$, 
Jack~O.~Burns$^8$, 
Chris~L.~Carilli$^9$, 
Renyue~Cen$^{10}$, 
Asantha~Cooray$^{11}$, 
Nickolay~Y.~Gnedin$^{12}$, 
Lincoln~J.~Greenhill$^2$,
Zoltan~Haiman$^{13}$, 
Jacqueline~N.~Hewitt$^{14}$,
Christopher~M.~Hirata$^7$,
Joseph~Lazio$^{15}$, 
Andrei~Mesinger$^{10}$, 
Piero~Madau$^{16}$,  
Miguel~F.~Morales$^{17}$,
S.~Peng~Oh$^{18}$, 
Jeffrey~B.~Peterson$^{19}$, 
Ylva~M.~Pihlstr{\" o}m$^{20}$, 
Max~Tegmark$^{14}$,
Hy~Trac$^2$, 
Oliver~Zahn$^6$, 
\& Matias~Zaldarriaga$^2$}

\affiliation{$^1$University of California, Los Angeles; sfurlane@astro.ucla.edu; (310) 206-4127
\\$^2$Harvard-Smithsonian Center for Astrophysics
\\$^3$University of Texas-Austin
\\$^4$University of Pennsylvania
\\$^5$Stanford University
\\$^6$University of California, Berkeley
\\$^7$California Institute of Technology
\\$^8$University of Colorado, Boulder
\\$^9$National Radio Astronomy Observatory
\\$^{10}$Princeton University
\\$^{11}$University of California, Irvine
\\$^{12}$Fermi National Accelerator Laboratory and University of Chicago
\\$^{13}$Columbia University
\\$^{14}$Massachusetts Institute of Technology
\\$^{15}$Naval Research Laboratory
\\$^{16}$University of California, Santa Cruz
\\$^{17}$University of Washington
\\$^{18}$University of California, Santa Barbara
\\$^{19}$Carnegie Mellon University, 
\\$^{20}$University of New Mexico
}

\maketitle

\vskip -0.3in

\noindent
Submitted for consideration by  the Astro2010 Decadal Survey Science Frontier Panel \\ \emph{Cosmology and Fundamental Physics}

\begin{center}
\includegraphics[width=3.8in]{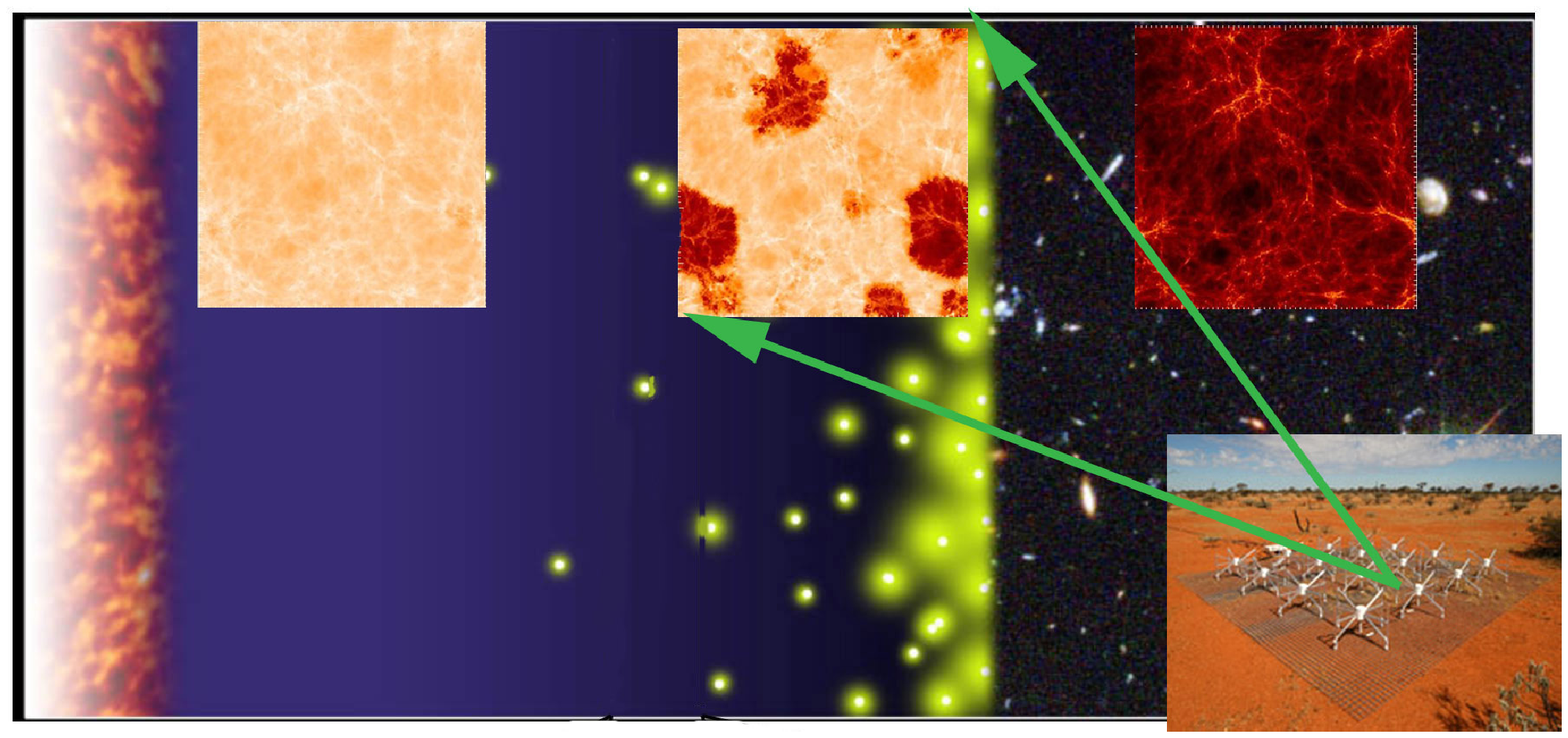}
\end{center} 

\vskip -0.2in

\noindent
The history of the Universe from recombination to the present, with example 21 cm signals and the Murchison Widefield Array overlaid.  Credit: S.~Furlanetto, J.~Lazio, and http://www.mwatelescope.org.

\newpage

\setcounter{page}{1}

\section{Introduction} \label{intro}

Measurements of our Universe's fundamental parameters have improved enormously over the past twenty years, thanks to probes as diverse as galaxy surveys \cite{percival07}, supernovae \cite{kowalski08}, and the cosmic microwave background (CMB) \cite{dunkley08}.  But answering the many questions opened by these studies requires new cosmological tools.  Here we describe the enormous potential of the 21 cm transition of neutral hydrogen, with which we can map the otherwise inaccessible cosmic ``dark ages" (at $6 < z < 50$, during and before the ``reionization" of intergalactic hydrogen).  This era includes nearly $60\%$ of the (in principle) observable volume of the Universe and contains an astonishing $\sim 3 \times 10^{16}$ independent measurements \cite{loeb04} -- a billion times more than in the CMB -- thanks to small-scale structure over such a large volume.  The \emph{potential} for improved measurements of the fundamental cosmological parameters is impressive, with the eventual possibility of, e.g., tightening constraints on our Universe's curvature by two orders of magnitude.  Over the next decade, we will take the first steps toward unlocking this potential and answer two key questions:  {\bf Does the standard cosmological model describe the Universe during the ``dark ages?"} {\bf How does the IGM evolve during this important time, ending with the reionization of hydrogen?}

\section{Scientific Context} \label{basic}

The 21 cm brightness temperature of an IGM gas parcel at a redshift $z$, relative to the cosmic microwave background, is \cite{madau97, furl06-review}
\bq
\dtb \approx 25\;\xhi(1+\delta) \, \left( { 1+z \over 10} \right)^{1/2}\, \left[1-\frac{T_{\gamma}(z)}{T_S}\right] \, \left[ \frac{H(z)/(1+z)}{\deriv v_\parallel/\deriv r_\parallel} \right] \mkel,
\label{eq:dtb}
\eq
where $\xhi$ is the neutral fraction, $\delta = \rho/\VEV{\rho}-1$ is the fractional IGM overdensity in units of the mean, $T_{\gamma}$ is the CMB temperature, $T_S$ is the spin (or excitation) temperature of this transition, $H(z)$ is the Hubble constant, and $\deriv v_\parallel/\deriv r_\parallel$ is the line-of-sight velocity gradient.

\begin{figure}
\includegraphics[width=4in]{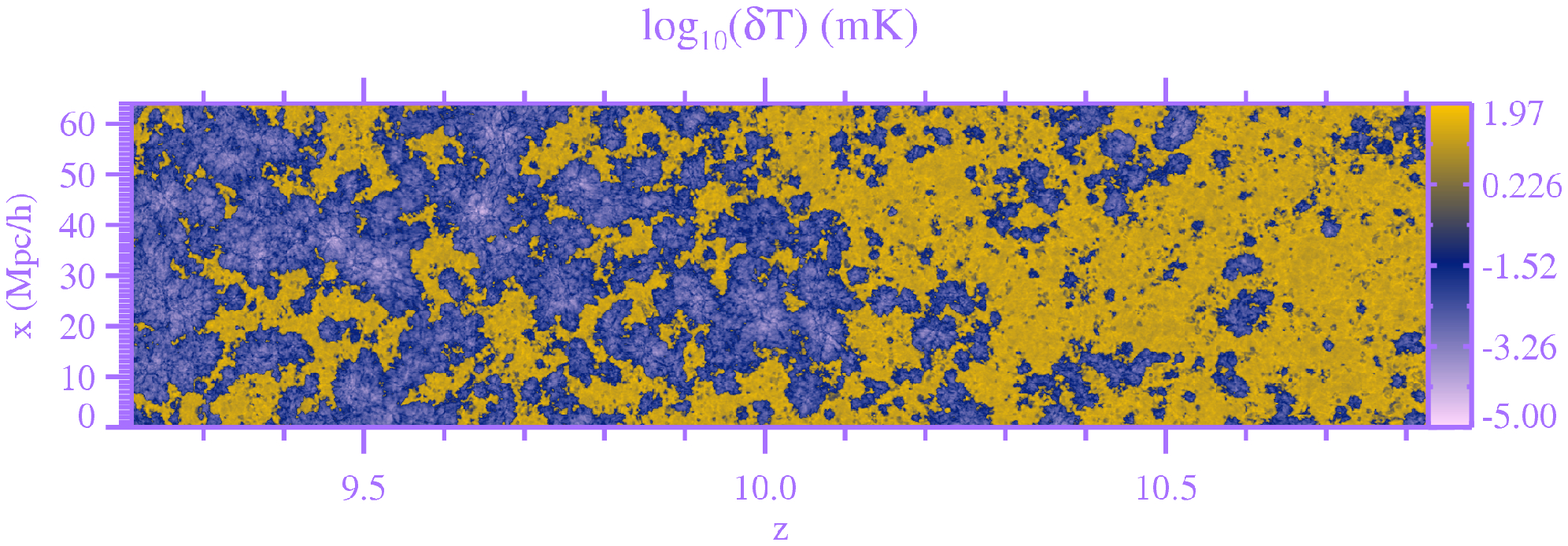}
\includegraphics[width=4in]{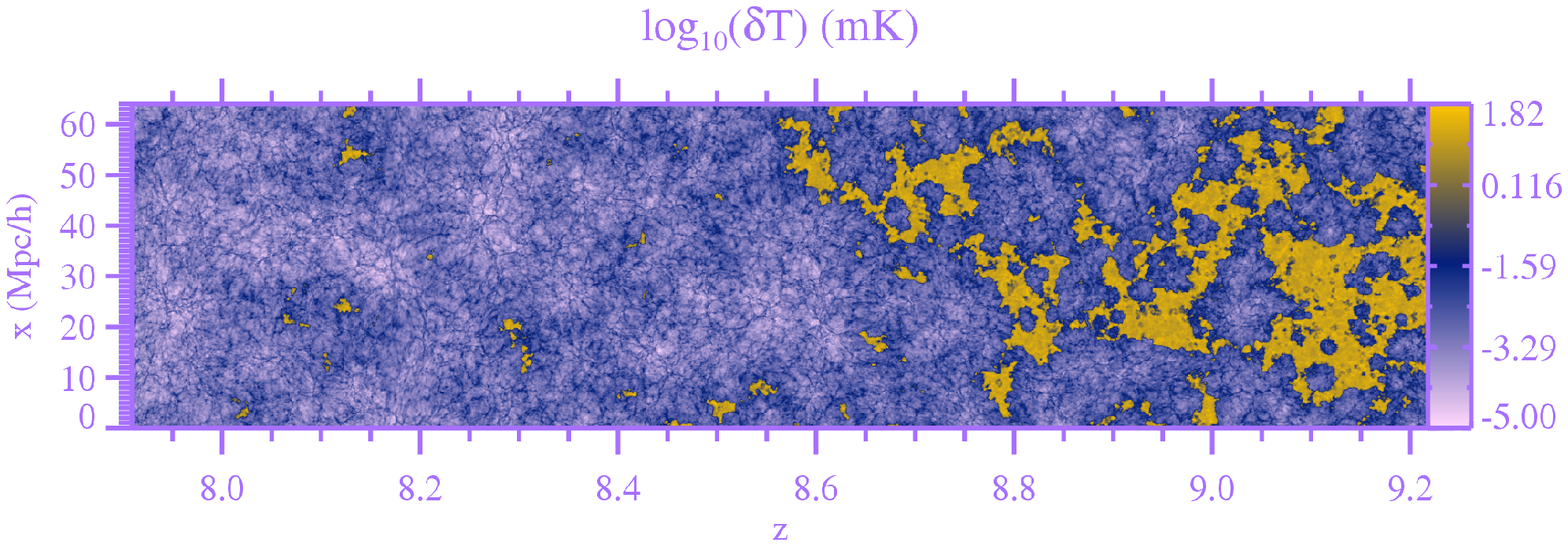}
\caption{Simulated maps of the 21 cm background during the early and late stages of a particular reionization scenario (top and bottom panels).  Purple regions are highly ionized; yellow regions are mostly neutral.  In the vertical direction, the slice subtends $\sim 35$' on the sky.  From \cite{shapiro08}.}
\label{fig:pics}
\end{figure}

All of these contributions contain unique cosmological information, some of which is illustrated by Fig.~\ref{fig:pics}.  The dependence on $\delta$ traces the development of the cosmic web \cite{scott90}, while the velocity factor sources line-of-sight ``redshift-space distortions" described below.  The other two factors depend strongly on the ambient radiation fields in the early Universe:  the ionizing background for $\xhi$ and a combination of the ultraviolet background (which mixes the 21 cm level populations through the Wouthuysen-Field effect \cite{wouthuysen52, field58}) and the X-ray background (which heats the gas \cite{chen04}) for $T_S$.  

\begin{figure}[htbp]
\begin{center}
\includegraphics[scale=0.35]{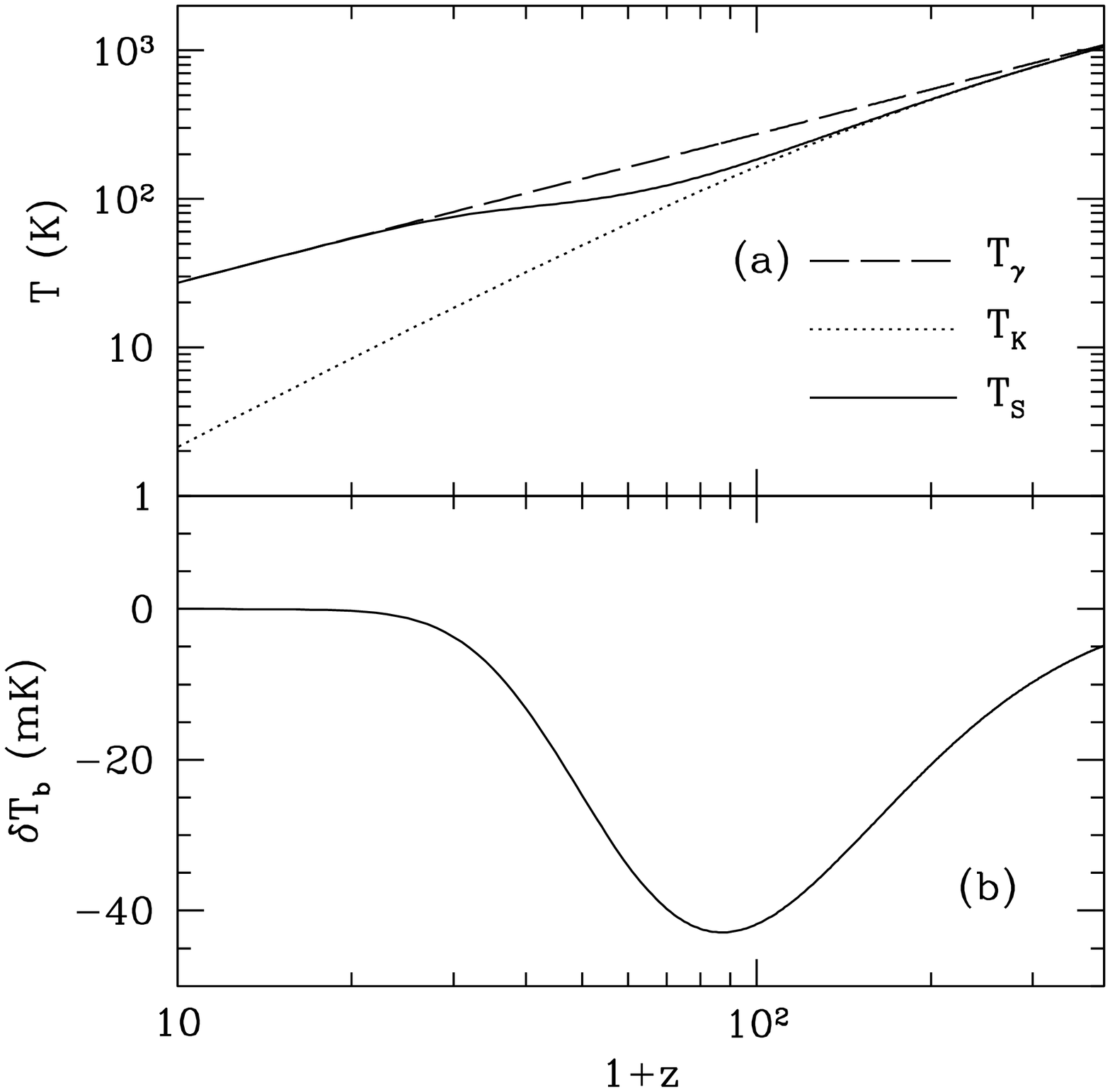}
\includegraphics[scale=0.35]{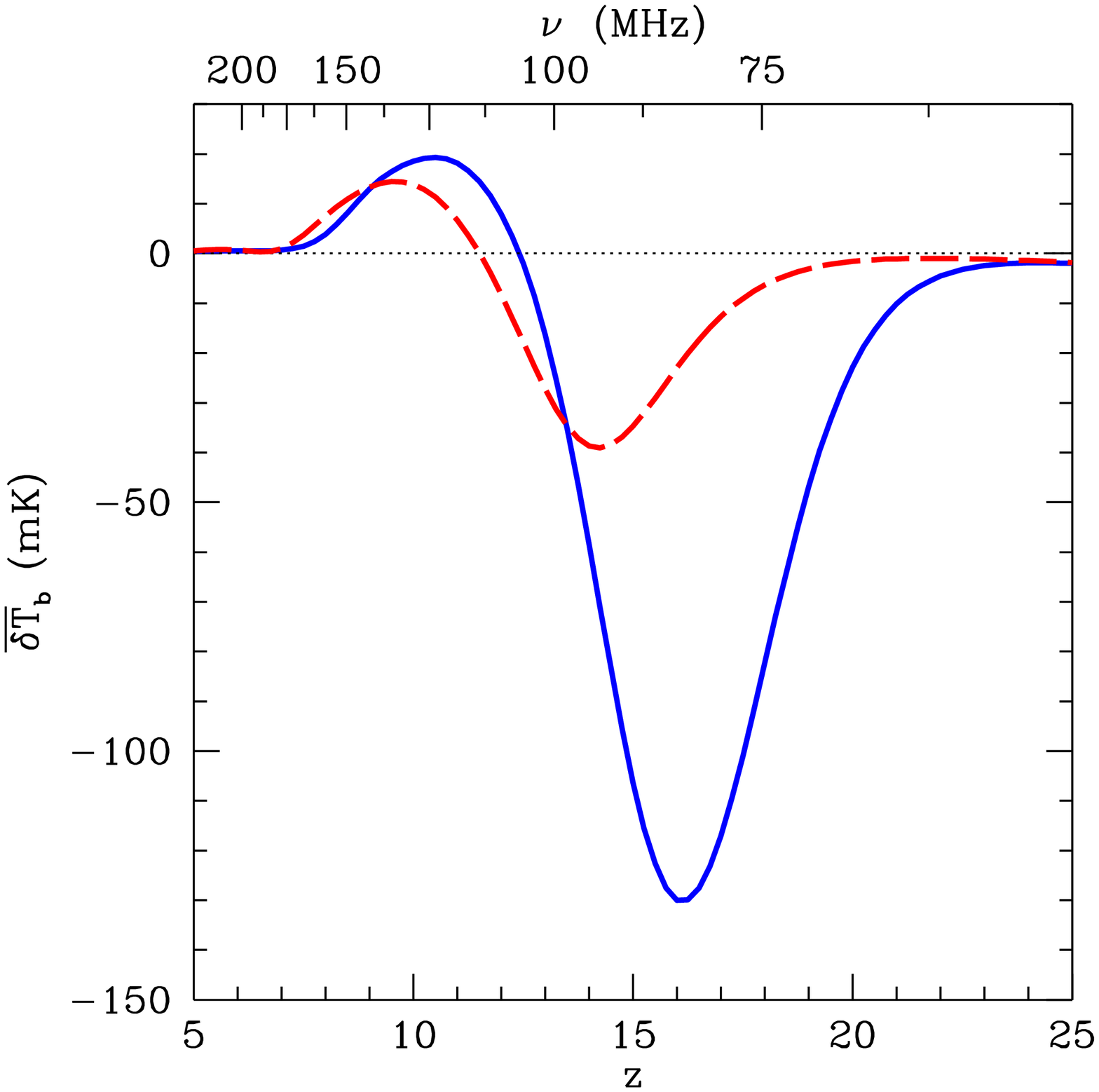}
\caption{\emph{Left:}  Thermal history of the IGM without star formation (\emph{top}) and the resulting sky-averaged 21 cm brightness temperature $\bdtb$ (\emph{bottom}).  \emph{Right:}  Fiducial histories of $\bdtb$ after star formation begins.  The solid blue (dashed red) curves use typical Population II stars (very massive Population III stars); both fix reionization to end at $z_r \approx 7$. From \cite{furl06-glob}. }
\label{fig:global}
\end{center}
\end{figure}

Fig.~\ref{fig:global} shows some example scenarios for the sky-averaged 21 cm brightness temperature.  The left panel ignores star formation:  in that case, we find 21 cm absorption at $z>50$ (because of strong collisional $T_S$ coupling) that fades at lower redshifts as the collision rate decreases.  In contrast, the right panel shows two models  of $\bdtb$ after galaxy formation begins (with normal and exotic stars), illustrating how the parameters of high-$z$ galaxy formation strongly affect the 21 cm background.  In both cases, ultraviolet photons from these stars first trigger strong absorption against the CMB \cite{madau97}.  Then, X-rays produced by the first supernovae or black holes heat the gas, turning the absorption into emission, which fades and eventually vanishes as ionizing photons destroy the intergalactic HI.  The 21 cm background thus depends strongly on unknown astrophysical parameters, and cosmological constraints will require a way to separate these ``contaminants."

\section{Key Questions} \label{cos}

We have so far described the 21 cm signal through the all-sky background, but of course the sky actually fluctuates strongly as individual IGM regions collapse gravitationally and become heated and ionized by luminous sources.  These fluctuations can ideally be imaged, but over the short term statistical measurements will be more powerful  (see \S \ref{mile} below).  Fortunately, as we discuss next (also see Fig.~\ref{fig:pk}), such statistics contain an enormous amount of information about the ``dark ages"  (also see the companion white paper ``Astrophysics from the Highly-Redshifted 21 cm Transition" for related questions about galaxy formation).
 
{\bf Does the standard cosmological model describe the Universe during the ``dark ages?"}  Perhaps the most important aspect of these measurements is the opening of an entirely new cosmological era to precision tests, to determine whether our standard model adequately describes this epoch.  The prospects for ``new" physics are difficult to quantify, but there are two clear opportunities to improve our understanding of the standard cosmology:

{\it What are the fundamental cosmological parameters of our Universe?}  In addition to opening the $z>6$ realm, the 21 cm background allows us to study the matter power spectrum on \emph{smaller} scales than any other -- stretching to $k > 100 \Mpcinv$, with much of this range even remaining linear.  With the enormous volumes available to 21 cm surveys, they can \emph{potentially} dramatically improve cosmological constraints on parameters such as the inflationary power spectrum, the neutrino mass, and the curvature of the Universe \cite{mcquinn06-param, bowman07, pritchard08-nu, mao08}.

The challenge will lie in separating astrophysical processes from the ``pure" cosmological information.  Three basic methods may work.  The first is robust modeling of the astrophysics \cite{mao08}.  Second, there may be an era in which astrophysical effects can be ignored (Fig.~\ref{fig:global}; certainly at $z >30$ before the first stars appear \cite{shapiro06} and plausibly just before reionization as well \cite{furl06-review}).  Third, we can take advantage of \emph{redshift-space distortions} of the 21 cm background, in which peculiar velocities change the mapping from frequency to radial distance, amplifying fluctuations along the line of sight (but not in the plane of the sky).  The degree of amplification depends on the angle cosine $\mu$ with the line of sight, and we can measure the distortions by breaking the power spectrum into its angular components (shown in the right panels of Fig.~\ref{fig:pk}).  Most importantly, the peculiar velocities depend almost only on gravity and so provide a purer view of the matter power spectrum \cite{barkana05-vel}.  

\begin{figure}[htbp]
\begin{center}
\includegraphics[scale=0.35]{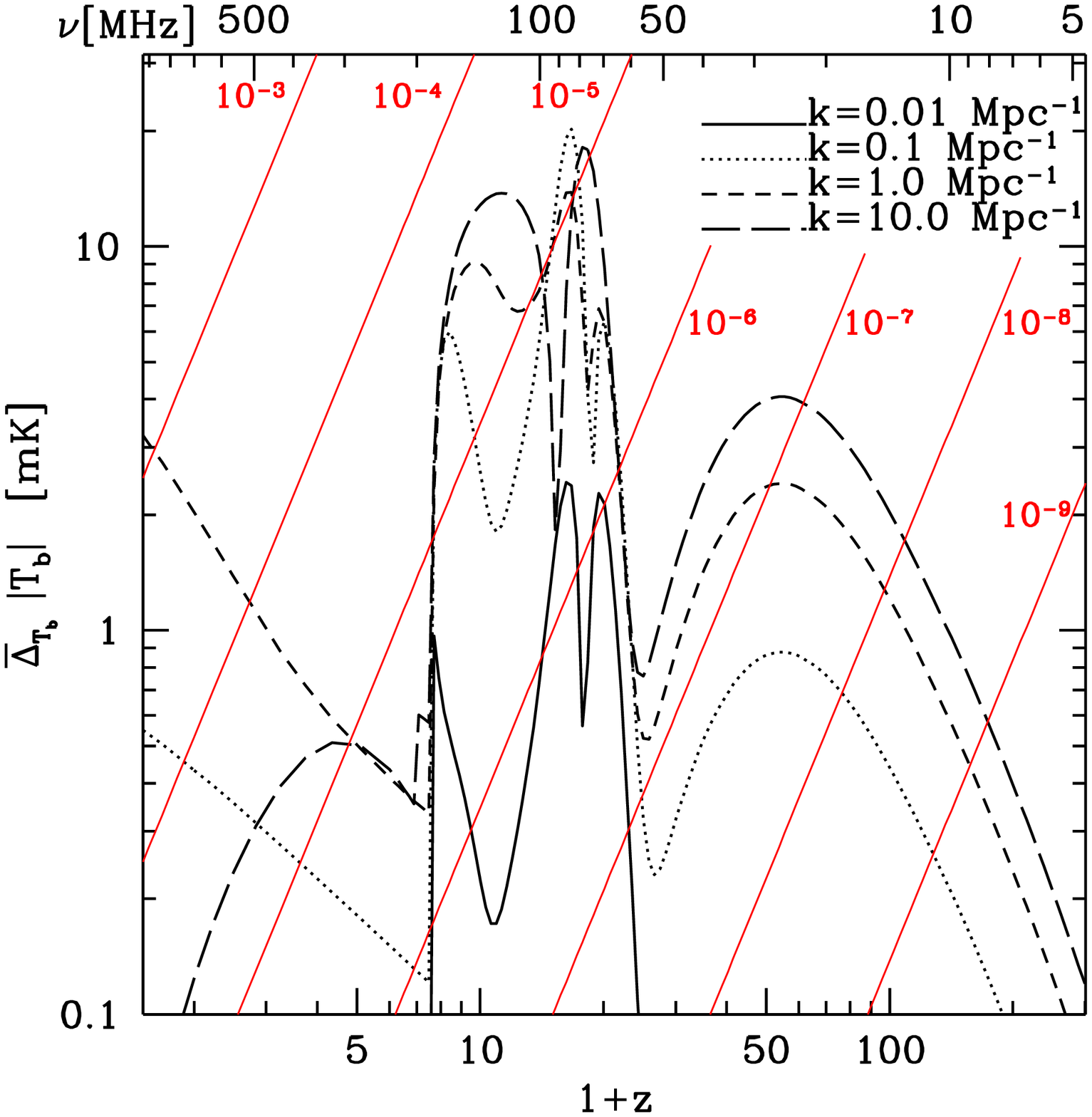}
\includegraphics[scale=0.35]{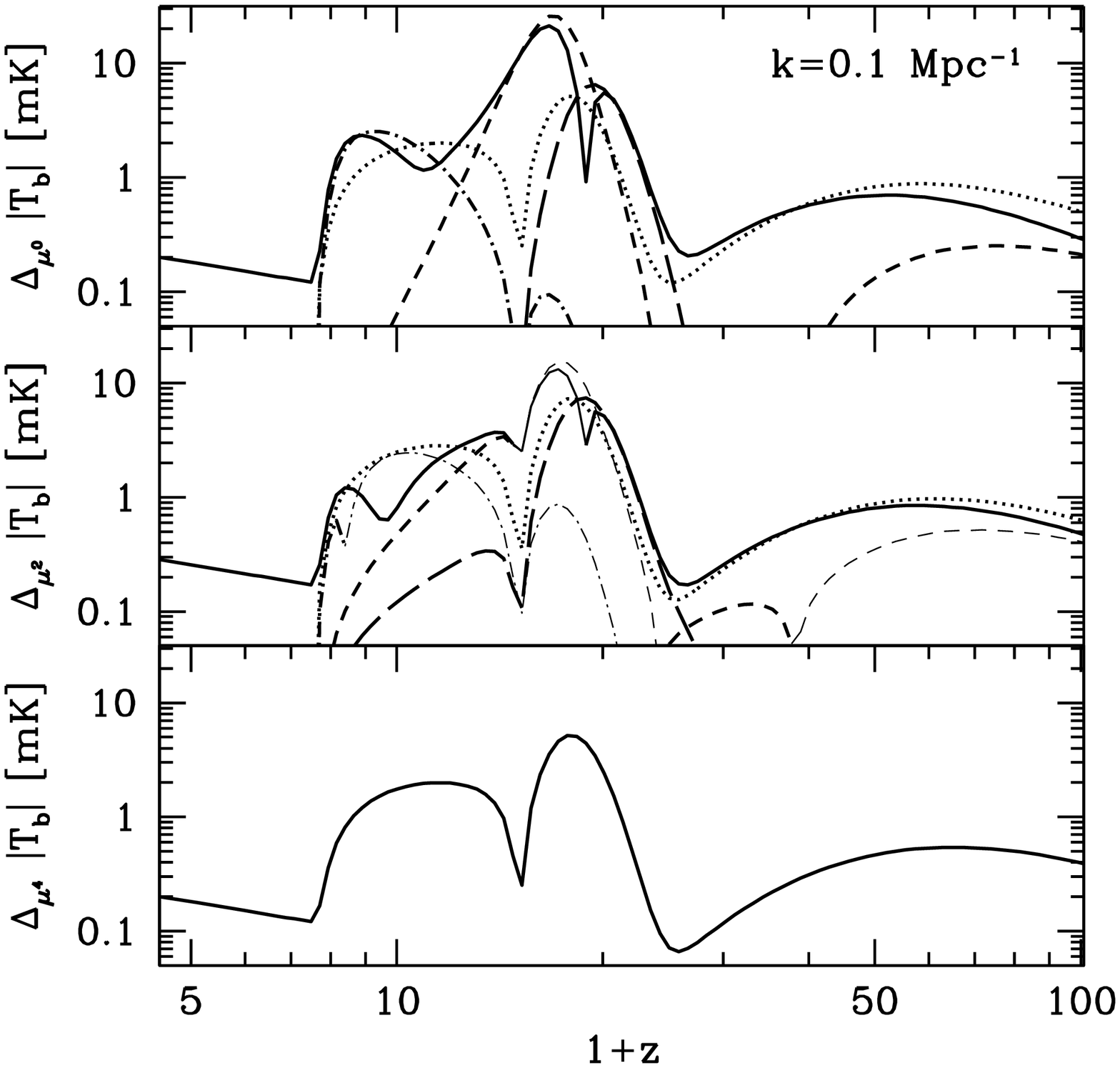}
\caption{\emph{Left:}  21 cm power spectrum fluctuations throughout the ``dark ages," at several different spatial wavenumbers $k$.  The red solid diagonal lines show the precision to which foregrounds must be removed in order to measure these fluctuations.  \emph{Right:}  21 cm fluctuations in the three angular components of the power spectrum when redshift space-distortions are included (solid curves), for $k=0.1 \Mpcinv$ (near the peak sensitivity of most arrays).  From top to bottom, these correspond to the isotropic, $\mu^2$, and $\mu^4$ components; the latter is quite insensitive to astrophysical ``contaminants."  The other curves show various contributions to the total signals.  From \cite{pritchard08}.}
\label{fig:pk}
\end{center}
\end{figure}

{\it What is the nature of dark matter?}  Before the first stars form, the 21 cm background depends only on simple physics -- recombination and linear perturbation theory in an expanding Universe \cite{loeb04}.  This makes it a sensitive probe of any exotic process affecting the thermal history of the IGM (much more so than the CMB), including dark matter decay or annihilation (even in standard dark matter models \cite{furl06-dm, myers08}), primordial black holes \cite{mack08}, and cosmic strings \cite{khatri08}.  These same processes also indirectly affect early galaxy formation and hence the 21 cm background at later times.

{\bf How did the IGM evolve during the ``dark ages?"}  21 cm experiments also provide the premier tool to study the evolution of the baryonic IGM and the beginning stages of galaxy formation -- especially reionization, the most dramatic event in the IGM's history.  In fact, the 21 cm background is the \emph{ideal} probe of reionization, which imprints strong fluctuations in it (see the middle panel in Fig.~\ref{fig:pics} and the large signal at $z \sim 10$ in Fig.~\ref{fig:pk}).  Its weak oscillator strength (in comparison to Ly$\alpha$) allows us to penetrate even extremely high redshifts.  We can also image it across the entire sky -- instead of only rare, isolated \lya forest lines of sight.  Moreover, unlike the CMB, it is a spectral line measurement, and we can distinguish different redshift slices to study the full history of the ``dark ages" -- extremely difficult even with a ``perfect" CMB measurement \cite{zald08-cmbpol}.  Finally, it directly samples the $ 95\%$ (or more) of the baryons that reside in the IGM. 

Observations have provided tantalizing hints about reionization, but even more unanswered questions \citep{fan06-review}.  For example, CMB observations imply that reionization completed by $z \sim 10$ \citep{dunkley08}, but quasar absorption spectra suggest that it may have continued until $z \sim 6$ \cite{fan06}, albeit both with substantial uncertainties.  Both the sky-averaged $\bdtb$ and the 21 cm power spectrum yield much more precise measures of $\bxhi(z)$ (see Fig.~\ref{fig:global} and the left panel of Fig.~\ref{fig:pk}; the fluctuations grow as ionized bubbles appear and then fade as the ionized regions fill the Universe).  The principal goal of first-generation experiments (now under construction) is to constrain this time evolution \cite{lidz08-constraint}.  But the 21 cm background contains much more detailed information on the ionizing sources, their interactions with the IGM, and their feedback mechanisms -- all of which have direct manifestations in the 21 cm power spectrum.  

At even higher redshifts, the 21 cm background is also sensitive to the growth of IGM structure (through the baryonic power spectrum \cite{barkana05-infall}) and the feedback processes that affect it.  In particular, the UV and X-ray backgrounds are responsible for heating the IGM as the \lya forest forms (even affecting its observable properties at $z<6$ \cite{hui03}).  They also trigger the strong increase in 21 cm fluctuations at $z \sim 20$ in Fig.~\ref{fig:pk}.  The 21 cm background provides the \emph{only} known method to study this ``pre-galactic" phase of structure formation.

\section{Milestones} \label{mile}

The ultimate goal of studying the 21 cm background is to make detailed maps of the IGM throughout the ``dark ages" and reionization, as in Fig.~\ref{fig:pics}.  The top axis of Fig.~\ref{fig:global} shows the observed frequency range for these measurements: well within the low-frequency radio regime.  Unfortunately, this is an extremely challenging band, because of terrestrial interference, ionospheric refraction, and (especially) other astrophysical sources (see \cite{furl06-review}).  In particular, the polarized Galactic synchrotron foreground has $T_{\rm sky} \sim 180 (\nu/180 \MHz)^{-2.6} \kel$, at least four orders of magnitude larger than the signal.  For an interferometer, the noise per resolution element (with an angular diameter $\Delta \theta$ and spanning a bandwidth $\Delta \nu$) is then \cite{furl06-review}
\begin{equation}
\Delta T_{\rm noise} \sim 20  \mkel \ \left( \frac{10^4 \sqm}{A_{\rm eff}} \right) \, \left( \frac{10'}{\Delta \theta} \right)^2 \, \left( \frac{1+z}{10} \right)^{4.6} \,  \left( \frac{{\rm MHz}}{\Delta \nu} \, \frac{100 \hr}{t_{\rm int}} \right)^{1/2},
\label{eq:if-sens}
\end{equation}
where $A_{\rm eff}$ is the effective collecting area and $t_{\rm int}$ is the integration time.  These angular and frequency scales correspond to $\sim 30 \Mpc$.  Here we outline the steps required to explore this era in detail, given the challenges implicit in this huge noise.

{\bf The all-sky signal:}  The global background illustrated in Fig.~\ref{fig:global} contains an extraordinary amount of cosmological information.  These measurements can easily beat down the noise with only a single dipole, so they may provide our first constraints at very high redshifts.  The challenge lies in calibration that is precise enough to extract the signal from instrumental artifacts and the bright foregrounds.  The EDGES experiment \cite{bowman08} has already set upper limits and hopes to measure this signal to $z \sim 20$ over the next decade.

\begin{figure}[htbp]
\begin{center}
\includegraphics[scale=0.35]{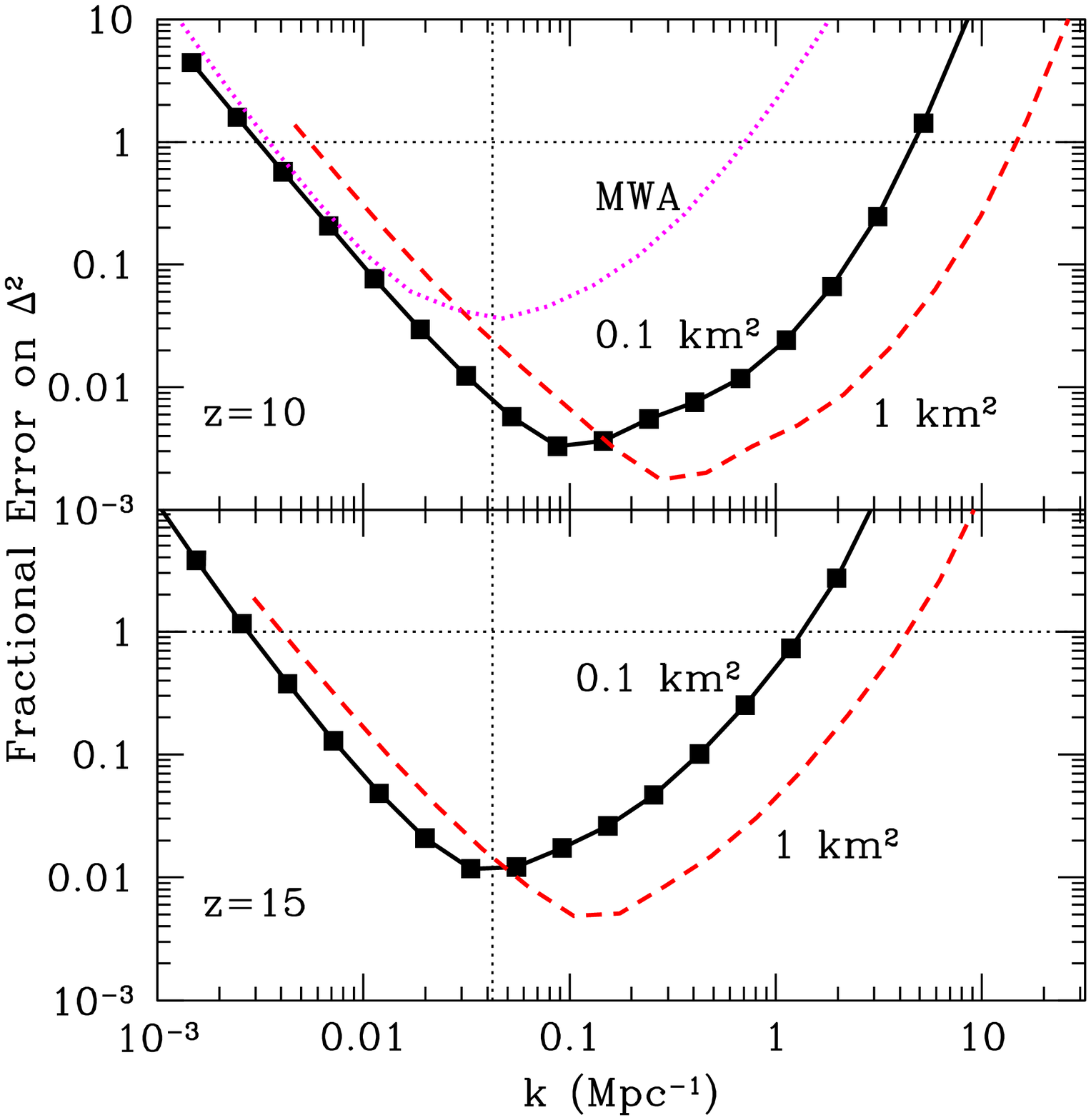}
\includegraphics[scale=0.35]{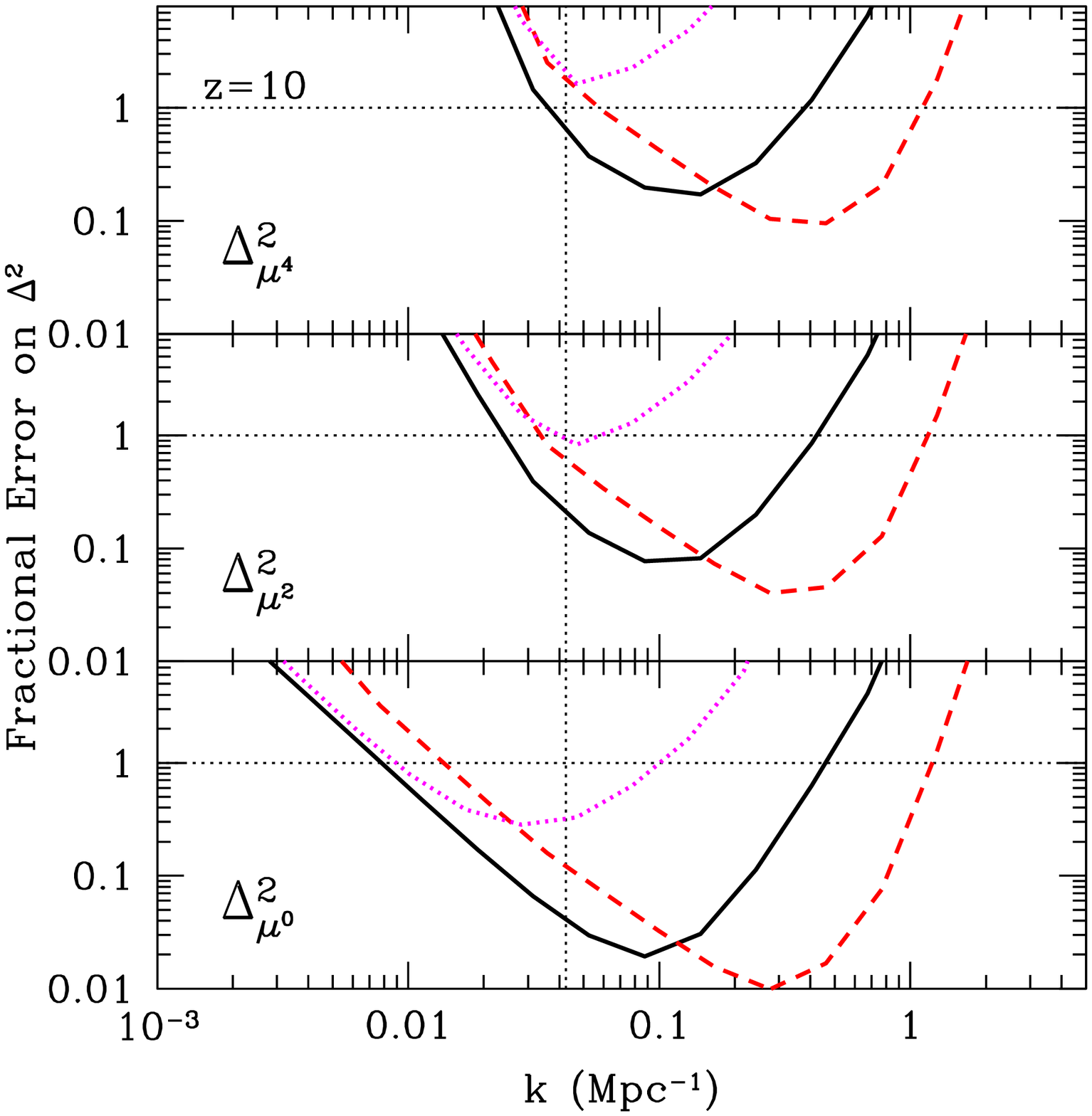}
\caption{\emph{Left panels:}  Sensitivity of some fiducial arrays to the spherically-averaged power spectrum at $z=10$ and $z=15$, (conservatively) assuming a fully neutral emitting IGM.  The solid and dashed curves take $A_{\rm eff}=0.1$ and 1 km$^2$, respectively; in both cases we take  4000 hr of observing split over four fields, a total bandwidth $B=8 \MHz$, $T_{\rm sky} = (471,\,1247) \kel$ at $z=(10,15)$, and $N=5000$ stations centered on a filled core and with an envelope out to $R_{\rm max}=3$~km.  The black squares show the location of the independent $k$ bins in each measurement.  The dotted curve shows the projected MWA sensitivity at $z=10$.  The vertical dotted line corresponds to the bandwidth; modes with smaller wavenumbers are compromised by foreground removal.  \emph{Right panels:} Sensitivity of the same arrays to angular components of the power spectrum at $z=10$.}
\label{fig:sensitivity}
\end{center}
\end{figure}

{\bf First-Generation Arrays:}  The arrays now operating or under construction, including the Murchison Widefield Array (MWA), the Precision Array to Probe the Epoch of Reionization (PAPER), LOFAR, the 21 Centimeter Array, and the Giant Metrewave Radio Telescope, have $A_{\rm eff} \sim 10^4 \sqm$ and so are limited to imaging only the most extreme ionized regions (such as those surrounding bright quasars).  Nevertheless,  these arrays have sufficiently large fields of view ($> 400^{\circ2}$) to make reasonably good statistical measurements \cite{mcquinn06-param, bowman07}.  Fig.~\ref{fig:sensitivity} shows the projected errors for the MWA at $z=10$.  It should detect fluctuations over a limited spatial dynamic range but only at $z < 12$, constraining reionization but offering no significant improvements to cosmological measurements unless strong deviations from the standard model exist during the ``dark ages."  (In the near term, lower-$z$ 21 cm surveys, operating at higher frequencies, may be more useful for cosmological parameter estimation \cite{chang08, visbal08}.)

{\bf Second-Generation Arrays:}  Fig.~\ref{fig:sensitivity} shows that larger telescopes, with $A_{\rm eff} \sim 10^5 \sqm$ (and large fields of view), will clearly be needed for precise measurements, and especially for the smaller scale information most useful to cosmology.  For example, after one year of observing we can expect improvements by factors of $(1.7,\, 2.5,\, 1.5,\, 3.0,\, 1.4,\, 2.7)$ in constraints on (respectively) the dark energy density, matter density, baryon density, neutrino mass, power spectrum spectral index, and its running, over those available from the \emph{Planck} CMB telescope on its own \cite{mcquinn06-param}, provided that the astrophysical ``contamination" can be cleaned.  

Fortunately, instruments in this class will also be able to measure some more advanced statistics.  The right panels in Fig.~\ref{fig:sensitivity} illustrate the sensitivity to the redshift-space distortions (with the $\mu^4$ component most useful for cosmology in the uppermost panel).  Although the limited sensitivity means that the velocity term on its own cannot provide high precision constraints, it will be extremely useful for breaking degeneracies between the many contributions to the signal \cite{barkana05-vel} -- to allow the constraints described above to be realized during and before reionization (especially with improved theoretical modeling \cite{mao08}).  

{\bf Imaging Arrays:}  At $A_{\rm eff} \sim 10^6 \sqm$, imaging on moderate scales becomes possible, and statistical constraints become exquisite even at high redshifts  (provided that the large field of view, not strictly necessary for imaging, is maintained; see Fig.~\ref{fig:sensitivity}).  Imaging greatly increases the likelihood of cleanly separating the cosmological information.  Under moderately optimistic assumptions, instruments of this scale could improve constraints on inflationary parameters, curvature, and neutrino mass by one or two orders of magnitude \cite{mao08}.

Plans for these later generations will evolve as we learn more about ``dark age" physics  and the experimental challenges ahead; for example, the Long Wavelength Array and other lower-frequency instruments will study the ionospheric calibration required to explore the high-$z$ regime ($z > 12$, or $\nu < 110 \MHz$) and help determine the relative utility of a terrestrial Square Kilometer Array or a far-side Lunar Radio Array.  At the same time, we must explore whether innovative new telescope designs more closely aligned with the observables, such as the FFT Telescope \cite{tegmark09}, can provide cost-effective improvements.

This roadmap, with accompanying efforts to improve theoretical modeling of the first galaxies and data analysis techniques will allow us to explore the major science questions of the first phases of structure formation and cosmology during the Universe's ``dark ages."

\newpage

\bibliographystyle{apj}
\bibliography{Ref_composite}

\begin{thebibliography}{33}
\expandafter\ifx\csname natexlab\endcsname\relax\def\natexlab#1{#1}\fi

\bibitem[{{Barkana} \& {Loeb}(2005{\natexlab{a}})}]{barkana05-vel}
{Barkana}, R., \& {Loeb},  A. 2005{\natexlab{a}}, Astrophys. J., 624, L65

\bibitem[{{Barkana} \& {Loeb}(2005{\natexlab{b}})}]{barkana05-infall}
---. 2005{\natexlab{b}}, \mnras, 363, L36

\bibitem[{{Bowman} {et~al.}(2007){Bowman}, {Morales}, \& {Hewitt}}]{bowman07}
{Bowman}, J.~D., {Morales}, M.~F., \& {Hewitt}, J.~N. 2007, Astrophys. J., 661, 1

\bibitem[{{Bowman} {et~al.}(2008){Bowman}, {Rogers}, \& {Hewitt}}]{bowman08}
{Bowman}, J.~D., {Rogers}, A.~E.~E., \& {Hewitt}, J.~N. 2008, Astrophys. J., 676, 1

\bibitem[{{Chang} {et~al.}(2008){Chang}, {Pen}, {Peterson}, \&
  {McDonald}}]{chang08}
{Chang}, T.-C., {Pen}, U.-L., {Peterson}, J.~B., \& {McDonald}, P. 2008,
  Physical Review Letters, 100, 091303

\bibitem[{{Chen} \& {Miralda-Escud{\' e}}(2004)}]{chen04}
{Chen}, X., \& {Miralda-Escud{\' e}}, J. 2004, Astrophys. J., 602, 1

\bibitem[{{Dunkley} {et~al.}(2008){Dunkley}, {Komatsu}, {Nolta}, {Spergel},
  {Larson}, {Hinshaw}, {Page}, {Bennett}, {Gold}, {Jarosik}, {Weiland},
  {Halpern}, {Hill}, {Kogut}, {Limon}, {Meyer}, {Tucker}, {Wollack}, \&
  {Wright}}]{dunkley08}
{Dunkley}, J. et al. 2008,
  Astrophys. J., submitted (arXiv.org/0803.0586 [astro-ph])

\bibitem[{{Fan} {et~al.}(2006{\natexlab{a}}){Fan}, {Carilli}, \&
  {Keating}}]{fan06-review}
{Fan}, X., {Carilli}, C.~L., \& {Keating}, B. 2006{\natexlab{a}}, \araa, 44,
  415

\bibitem[{{Fan} {et~al.}(2006{\natexlab{b}}){Fan}, {Strauss}, {Becker},
  {White}, {Gunn}, {Knapp}, {Richards}, {Schneider}, {Brinkmann}, \&
  {Fukugita}}]{fan06}
{Fan}, X. et al. 2006{\natexlab{b}}, Astron. J., 132, 117

\bibitem[{{Field}(1958)}]{field58}
{Field}, G.~B. 1958, Proceedings of the Institute of Radio Engineers, 46, 240

\bibitem[{{Furlanetto}(2006)}]{furl06-glob}
{Furlanetto}, S.~R. 2006, \mnras, 371, 867

\bibitem[{{Furlanetto} {et~al.}(2006{\natexlab{a}}){Furlanetto}, {Oh}, \&
  {Briggs}}]{furl06-review}
{Furlanetto}, S.~R., {Oh}, S.~P., \& {Briggs}, F.~H. 2006{\natexlab{a}},
  \physrep, 433, 181

\bibitem[{{Furlanetto} {et~al.}(2006{\natexlab{b}}){Furlanetto}, {Oh}, \&
  {Pierpaoli}}]{furl06-dm}
{Furlanetto}, S.~R., {Oh}, S.~P., \& {Pierpaoli}, E. 2006{\natexlab{b}}, Phys. Rev. D,
  74, 103502

\bibitem[{{Hui} \& {Haiman}(2003)}]{hui03}
{Hui}, L., \& {Haiman}, Z. 2003, Astrophys. J., 596, 9

\bibitem[{{Khatri} \& {Wandelt}(2008)}]{khatri08}
{Khatri}, R., \& {Wandelt}, B.~D. 2008, Physical Review Letters, 100, 091302

\bibitem[{{Kowalski} {et~al.}(2008)}]{kowalski08}
{Kowalski}, M., {et~al.} 2008, Astrophys. J., 686, 749

\bibitem[{{Lidz} {et~al.}(2008){Lidz}, {Zahn}, {McQuinn}, {Zaldarriaga}, \&
  {Hernquist}}]{lidz08-constraint}
{Lidz}, A., {Zahn}, O., {McQuinn}, M., {Zaldarriaga}, M., \& {Hernquist}, L.
  2008, Astrophys. J., 680, 962

\bibitem[{{Loeb} \& {Zaldarriaga}(2004)}]{loeb04}
{Loeb}, A., \& {Zaldarriaga}, M. 2004, Physical Review Letters, 92, 211301

\bibitem[{{Mack} \& {Wesley}(2008)}]{mack08}
{Mack}, K.~J., \& {Wesley}, D.~H. 2008, submitted to \mnras \
  (arXiv.org/0805.1531 [astro-ph])

\bibitem[{{Madau} {et~al.}(1997){Madau}, {Meiksin}, \& {Rees}}]{madau97}
{Madau}, P., {Meiksin}, A., \& {Rees}, M.~J. 1997, Astrophys. J., 475, 429

\bibitem[{{Mao} {et~al.}(2008){Mao}, {Tegmark}, {McQuinn}, {Zaldarriaga}, \&
  {Zahn}}]{mao08}
{Mao}, Y., {Tegmark}, M., {McQuinn}, M., {Zaldarriaga}, M., \& {Zahn}, O. 2008,
  Phys. Rev. D, 78, 023529

\bibitem[{{McQuinn} {et~al.}(2006){McQuinn}, {Zahn}, {Zaldarriaga},
  {Hernquist}, \& {Furlanetto}}]{mcquinn06-param}
{McQuinn}, M., {Zahn}, O., {Zaldarriaga}, M., {Hernquist}, L., \& {Furlanetto},
  S.~R. 2006, Astrophys. J., 653, 815

\bibitem[{{Myers} \& {Nusser}(2008)}]{myers08}
{Myers}, Z., \& {Nusser}, A. 2008, \mnras, 384, 727

\bibitem[{{Percival} {et~al.}(2007){Percival}, {Cole}, {Eisenstein}, {Nichol},
  {Peacock}, {Pope}, \& {Szalay}}]{percival07}
{Percival}, W.~J. et al. 2007, \mnras, 381, 1053

\bibitem[{{Pritchard} \& {Loeb}(2008)}]{pritchard08}
{Pritchard}, J.~R., \& {Loeb}, A. 2008, Phys. Rev. D, 78, 103511

\bibitem[{{Pritchard} \& {Pierpaoli}(2008)}]{pritchard08-nu}
{Pritchard}, J.~R., \& {Pierpaoli}, E. 2008, Phys. Rev. D, 78, 065009

\bibitem[{{Scott} \& {Rees}(1990)}]{scott90}
{Scott}, D., \& {Rees}, M.~J. 1990, \mnras, 247, 510

\bibitem[{{Shapiro} {et~al.}(2006){Shapiro}, {Ahn}, {Alvarez}, {Iliev},
  {Martel}, \& {Ryu}}]{shapiro06}
{Shapiro}, P.~R. et al. 2006, Astrophys. J., 646, 681

\bibitem[{{Shapiro} {et~al.}(2008){Shapiro}, {Iliev}, {Mellema}, {Pen}, \&
  {Merz}}]{shapiro08}
{Shapiro}, P.~R., {Iliev}, I.~T., {Mellema}, G., {Pen}, U.-L., \& {Merz}, H.
  2008, in American Institute of Physics Conference Series, Vol. 1035, The
  Evolution of Galaxies Through the Neutral Hydrogen Window, ed. R.~{Minchin}
  \& E.~{Momjian}, 68--74

\bibitem[{{Tegmark} \& {Zaldarriaga}(2009)}]{tegmark09}
{Tegmark}, M., \& {Zaldarriaga}, M. 2009, Phys. Rev. D, submitted (arXiv.org/0805.4414
  [astro-ph])

\bibitem[{{Visbal} {et~al.}(2008){Visbal}, {Loeb}, \& {Wyithe}}]{visbal08}
{Visbal}, E., {Loeb}, A., \& {Wyithe}, S. 2008, Phys. Rev. D, submitted
  (arXiv.org/0812.0419 [astro-ph])

\bibitem[{{Wouthuysen}(1952)}]{wouthuysen52}
{Wouthuysen}, S.~A. 1952, Astron. J., 57, 31

\bibitem[{{Zaldarriaga} {et~al.}(2008){Zaldarriaga}, {Colombo}, {Komatsu},
  {Lidz}, {Mortonson}, {Oh}, {Pierpaoli}, {Verde}, \& {Zahn}}]{zald08-cmbpol}
{Zaldarriaga}, M. et al. 2008, CMBPol Mission
  Concept Study, (arXiv.org/0811.3918 [astro-ph])

\end{thebibliography}

\end{document}